\documentclass[runningheads]{llncs}
\newcommand\blfootnote[1]{%
  \begingroup
  \renewcommand\thefootnote{}\footnote{#1}%
  \addtocounter{footnote}{-1}%
  \endgroup
}

\usepackage{graphicx}
\usepackage{tabularray}
\usepackage{xcolor}
\usepackage{hyperref}
\begin{document}
\title{Learning Effect of Lay People in Gesture-Based Locomotion in Virtual Reality}
\titlerunning{Learning Effect of Lay People in Gesture-Based Locomotion...}
%
\author{Alexander Schäfer\inst{1}\orcidID{0000-0003-2356-5285} \and
Gerd Reis\inst{2}\and
Didier Stricker\inst{1,2}}
\authorrunning{Alexander Schäfer et al.}
%
\institute{TU Kaiserslautern, Gottlieb-Daimler-Strasse Building 47,
67663 Kaiserslautern, Germany 
\email{Alexander.Schaefer@dfki.de}\and
German Research Center for Artificial Intelligence, Trippstadterstr. 122, 67663 Kaiserslautern, Germany
\email{\{Gerd.Reis,Didier.Stricker\}@dfki.de}\\
}
\maketitle              
\begin{abstract}
Locomotion in Virtual Reality (VR) is an important part of
VR applications. Many scientists are enriching the community with different variations that enable locomotion in VR. Some of the most promising methods are gesture-based and do not require additional  handheld hardware. Recent work focused mostly on user preference and performance of the different locomotion techniques. This ignores the learning effect that users go through while new methods are being explored. In this work, it is investigated whether and how quickly users can adapt to a hand gesture-based locomotion system in VR. Four different locomotion techniques are implemented and tested by participants. The goal of this paper is twofold: First, it aims to encourage researchers to consider the learning effect in their studies. Second, this study aims to provide insight into the learning effect of users in gesture-based systems.

\blfootnote{This version of the article has been accepted for publication, after peer review and is subject to Springer Nature’s AM terms of use, but is not the Version of Record and does not reflect post-acceptance improvements, or any corrections. The Version of Record is available online at: \url{https://doi.org/10.1007/978-3-031-05939-1_25}.}

\end{abstract}

\section{Introduction}
\label{sec:introduction}
Intuitive locomotion is an essential part of VR research and its applications. Usually a controller is used for virtual locomotion \cite{frommel2017effects,weissker2018spatial,rantala2021comparison}, but more recent work uses other techniques such as vision-based tracking \cite{schafer2021controlling} or sensors that are attached to the body \cite{sarupuri2017triggerwalking,pai2017armswing}. The controller-based methods are extensively researched and already in commercial use.  Gesture-based locomotion is getting more attention lately and requires more research to find out which methods are adequate. Many studies focus entirely on subjective user preference in researching and developing such locomotion systems. Moreover, studies related to locomotion in VR usually use a single "ease-to-learn" question for participants to find out if the technique is easy to learn. This article addresses another important factor that is often neglected. Namely, how quickly people can learn such a system. Complicated methods lead to longer times to achieve objectives and ultimately to user frustration. Therefore, attention should be paid to whether or not users can quickly adapt to implemented techniques. \par 

To investigate the learning effect, a user study with 21 participants was conducted. Four different locomotion techniques based on hand gestures were implemented and tested by each participant. People with minor and no background in VR were recruited in order to remove additional bias or knowledge regarding VR systems. This article focuses on the learning effect that can be observed among these lay people. More concrete, this work answers the following research questions: \par

\begin{description}
\item \emph{Do lay people significantly improve their task completion time when using hand gesture based locomotion during a second session compared to the first?} \\
\item \emph{How much do lay people adapt to the limitations of a hand tracking device after a first trial session?}\\
\item \emph{Will lay people significantly improve their efficiency (less number of teleportations) when using hand gesture based locomotion during a second trial session?}
\end{description}

\par
This article is organized as follows: In \autoref{sec:background} relevant articles about the learning effect in VR locomotion are provided. The locomotion techniques used for the experiment are introduced in  \autoref{sec:locomotiontechniques}. In \autoref{sec:experiment} the experiment is described and \autoref{sec:answering} the research questions are answered. Finally, the results are discussed in \autoref{sec:discussion} and concluded in 
\autoref{sec:conclusion}.

\section{Background and Related Work}
\label{sec:background}
To the best of our knowledge, no paper was published that is focused on investigating the learning effect of lay people for hand gesture based VR locomotion. However, some work mentions learning effect during their studies. \par 
Zhao et al. \cite{zhao2021comparing} investigated different techniques to control locomotion speed. The gestures included Finger Distance, Finger Number, and Finger Tapping. Users were asked within a questionnaire to subjectively evaluate the ease-to-learn. According to the results, users found the proposed techniques easy to learn. However, no quantitative analysis was performed to gain insights into the learning effect of participants. \par 
Zielasko et al. \cite{zielasko2016evaluation} implemented and evaluated five different hands-free navigation methods for VR. The techniques included Walking In Place, Accelerator Pedal, Leaning, Shake Your Head, and Gamepad. Using a questionnaire, the authors came to the conclusion that the introduced techniques are easy to learn. \par 
Kitson et al. \cite{kitson2015navichair} introduced NaviChair, a chair based locomotion technique for virtual environments. Users were required to move within the chair to get different locomotion effects. The authors compared this technique with a technique based on a joystick. During exit interviews, it was revealed that the joystick variant is preferred by users because it was more accurate and easier to learn. \par 
The work of Keil et al. \cite{keil2021effects} used VR locomotion techniques to measure users' learning effect in distance estimations. The authors found a significant decrease in distance estimation errors after a subsequent task. \par 
This paper does not rely on a subjective questionnaire to answer whether techniques are easy to learn. Instead, gathered data was quantitatively analysed to measure the learning effect of users. 

\section{Locomotion Techniques}
\label{sec:locomotiontechniques}
\begin{figure}[h]
\includegraphics[width=\textwidth]{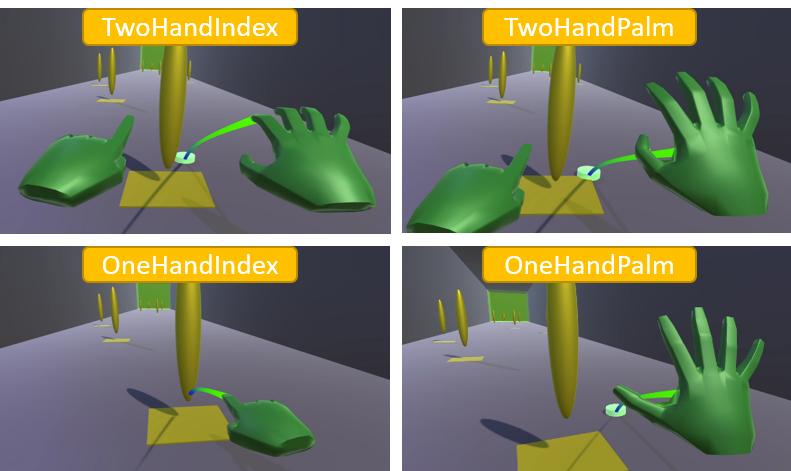}
\caption{The four locomotion techniques used for experiments. The two-handed techniques use one hand for choosing a destination and the other hand for activating the teleport via gestures. The one-handed techniques use one hand to choose the destination and the teleport is activated if the hand is kept still.} \label{fig1}
\label{fig:fourtechniques}
\end{figure}

In this section the locomotion techniques used for the experiment are introduced. Four different locomotion techniques, two two-handed and two one-handed, were implemented. All techniques are hand gesture-based and do not require controller or other handheld hardware. For details of the implementation of the proposed techniques, the reader is referred to the work of Schäfer et al. \cite{schafer2021controlling}. In the following text these four techniques are referenced as TwoHandIndex, TwoHandPalm, OneHandIndex, and OneHandPalm. These techniques are depicted in Figure \ref{fig:fourtechniques}.

\paragraph{TwoHandIndex} 
This method requires the user to use both hands for moving through a virtual environment. The index finger on the right hand is used to choose a teleport destination. If the left hand performs a certain gesture, teleportation to the chosen destination is performed. The gesture consists of bending and straightening the index finger, similar to a pointing gesture.

\paragraph{TwoHandPalm}
This method is similar to the TwoHandIndex technique. However, instead of using the index finger for navigation, the palm is used.

\paragraph{OneHandIndex}
Only one hand for navigation as well as performing teleportation is used for this technique. When a pointing gesture is found, a ray emanating from the tip of the index finger is activated. After the ray is visible, teleportation will be performed after the hand is held still for 1.5 s. 

\paragraph{OneHandedPalm}
Similar to OneHandIndex, only one hand is used for locomotion. Instead of a pointing gesture, this technique uses the palm for navigation. As soon as the user opens his hand, a ray emanates from the centre of the palm. The teleportation is triggered after the hand is held still for 1.5 s.

\section{Experiment}
\label{sec:experiment}
\subsection{Objectives}
The objective of the experiment was to answer the question whether users are able to quickly learn hand gesture based locomotion. More precise, the research questions mentioned in \autoref{sec:introduction} should be answered. Four techniques to control locomotion in VR with hand gestures were implemented and performed by users. Task completion time was used to measure the efficiency of the user with different techniques. Additionally, the number of teleportations required to reach the goal and the number of times the hand tracking failed was collected to measure improvements by the users.

\subsection{Participants}
\label{sec:participants}
The study was performed with 21 volunteers (15 Male, 6 Female). The subjects' age ranged between 25 and 60 years ($M$ = 35.4, $Median$ = 31). A 5-point Likert-scale (1 denotes less knowledge and 5 expert knowledge) was used as a preliminary questionnaire to assess the experience of participants within the relevant subject areas. 81 \% of participants answered that they have good experience in software and computer (they answered with 4 or 5). Regarding VR experience, 86 \% of users have never worn a VR-HMD before and the remaining 14 \% used a VR headset at least once.

\subsection{Apparatus}
\label{sec:apparatus}
The experiment was performed by using a gaming notebook. Hand tracking was provided by the Leap Motion Controller. Samsung Odyssey+ was used as the VR-HMD.
\subsection{Experimental Task}
\label{sec:task}
\begin{figure}[h]
\includegraphics[width=\textwidth]{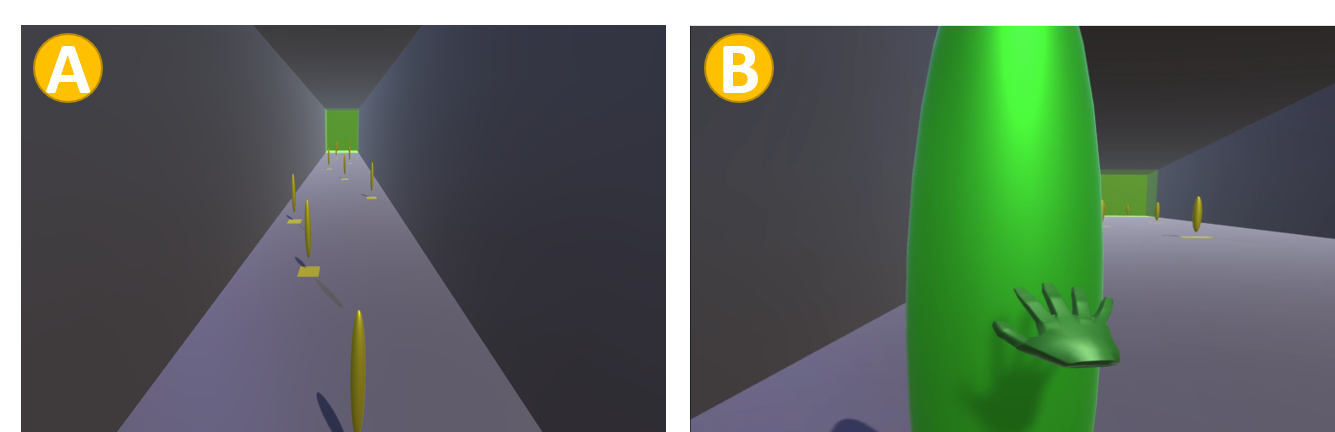}
\caption{(\textbf{A}) Overview of the virtual environment. 10 pillars are placed inside a large corridor. (\textbf{B}) Participant touching a pillar and changing its color to green.} \label{fig1}
\label{fig:experimentalTask}
\end{figure}

Participants had to touch ten pillar-like objects in a virtual environment. These objects were placed at fixed positions in a large corridor (10 m high, 10 m broad and 100 m long) with no obstacles other than the touchable pillar-like objects as depicted Figure \ref{fig:experimentalTask}A. \par 
Participants were required to move towards the pillars with the locomotion techniques proposed in \autoref{sec:locomotiontechniques}. Once the user was close enough, a pillar could be touched and its color changed to indicate it was touched (\autoref{fig:experimentalTask}B). The task was completed once the participant touched all 10 pillars in the VE. The task was repeated for each technique. The experiment was conducted in seating position.

\subsection{Procedure}
\label{sec:procedure}
The experiment had a within-subjects design. Each trial session was performed individually with the 21 subjects. The experiment was split into two phases: Learning phase and evaluation phase. In both phases, the participants had to move through the environment and perform the experimental task. In the learning, as well as the evaluation phase, all four techniques were used by the participants. Both phases are identical in terms of task and VE. The difference between the two phases is the order of techniques, which was counterbalanced between participants in the evaluation phase. In the learning phase, participants have performed each technique for the first time. Therefore, data is collected where users performed each technique for the first and the second time. A trial session lasted about 43 minutes.

\section{Answering the Research Questions}
\label{sec:answering}
Data was collected during the experiment to answer the research questions. No questionnaire regarding learning experience was given to the participants and our findings are based solely on quantitative results. Task completion time measured the time a participant required to complete the task. More precisely, it represents the time from touching the first pillar to the last inside the VE. Additionally, the number of times the hand tracking was lost was collected. This value represents the tracking failures of the chosen device. The tracking failed if the participant moved a hand out of the reliable tracking range of the device. For two-handed locomotion techniques, both hands had to be tracked by the sensor and for one-handed techniques only the dominant hand had to be tracked. This measure is particularly useful to measure the adaption of the user to overcome the limitations of the tracking device. Furthermore, the number of teleportations a user required to reach the goal was collected. The raw values such as mean, median, standard deviation, minimum, and maximum are shown in Table \ref{table:quantitativeTable}. One-way ANOVA was used for statistical analysis and throughout the paper significance at the 0.05 level is reported.

\definecolor{TableBlue}{HTML}{1b98e0}
\definecolor{TableYellow}{HTML}{FFDB00}
\definecolor{TableGreen}{HTML}{80FF00}
\begin{longtblr}[
  caption = {Raw data collected during the experiment. Task completion time, number of times the hand tracking was lost during a session, and number of teleportations required to complete the task is shown.},
  label = {table:quantitativeTable},
]{
  colspec = {|X[4]|r|r|r|r|r|},
  rowhead = 1,
  hlines,
  row{even} = {gray9},
  row{2} = {TableBlue},
  row{11} = {TableYellow},
  row{20} = {TableGreen}
} 
\textbf{Task} & \textbf{Mean} & \textbf{MDN} & \textbf{SD} & \textbf{MIN} & \textbf{MAX} \\ 
\multicolumn{6}{c}{\textbf{Task Completion Time}} \\ 
M1 - Learning           & 33.19     & 31    & 18.51     & 17    & 102 \\ 
M2 - Learning           & 25.61     & 23    & 8.78      & 15    & 47 \\ 
M3 - Learning           & 23.00     & 22    & 4.38      & 18    & 36 \\ 
M4 - Learning           & 22.19     & 20    & 5.87      & 17    & 41 \\ 
M1 - Evaluation         & 25.76     & 24    & 4.21      & 18    & 37 \\ 
M2 - Evaluation         & 25.52     & 23    & 8.07      & 16    & 48 \\ 
M3  - Evaluation        & 23.04     & 22    & 5.21      & 18    & 44 \\ 
M4 - Evaluation         & 22.33     & 21    & 4.62      & 17    & 35 \\ 
\multicolumn{6}{c}{\textbf{Hand Tracking Failures}} \\ 
M1 - Learning & 26.14 & 22 & 16.21 & 8 & 78 \\ 
M2 - Learning & 19.57 & 15 &10.68 & 7 & 46 \\ 
M3 - Learning & 12.14 & 11 &8.48 & 2 & 42 \\ 
M4 - Learning & 6.80 & 6 &3.84 & 2 & 14 \\ 
M1 - Evaluation   & 12.90 & 11 & 5.21 & 7 & 29 \\ 
M2 - Evaluation   & 13.28 & 11 & 5.55 & 6 & 28 \\ 
M3 - Evaluation   & 9.19 & 7 & 4.40 & 3 & 18 \\ 
M4 - Evaluation   & 8.95 & 7 & 6.0 & 3 & 29 \\ 
\multicolumn{6}{c}{\textbf{Number of Teleportations}} \\ 
M1 - Learning          & 33.19     & 31    & 18.51     & 17    & 102 \\ 
M2 - Learning          & 25.61     & 23    & 8.78      & 15    & 47 \\ 
M3 - Learning          & 23.00     & 22    & 4.38      & 18    & 36 \\ 
M4 - Learning          & 22.19     & 20    & 5.87      & 17    & 41 \\ 
M1 - Evaluation    & 25.76     & 24    & 4.21      & 18    & 37 \\ 
M2 - Evaluation     & 25.52     & 23    & 8.07      & 16    & 48 \\ 
M3  - Evaluation    & 23.04     & 22    & 5.21      & 18    & 44 \\ 
M4 - Evaluation    & 22.33     & 21    & 4.62      & 17    & 35 \\ 
\end{longtblr}

\subsection{Do lay people significantly improve their task completion time when using hand gesture based locomotion during a second session compared to the first?}
\begin{figure}[h]
\includegraphics[width=\textwidth]{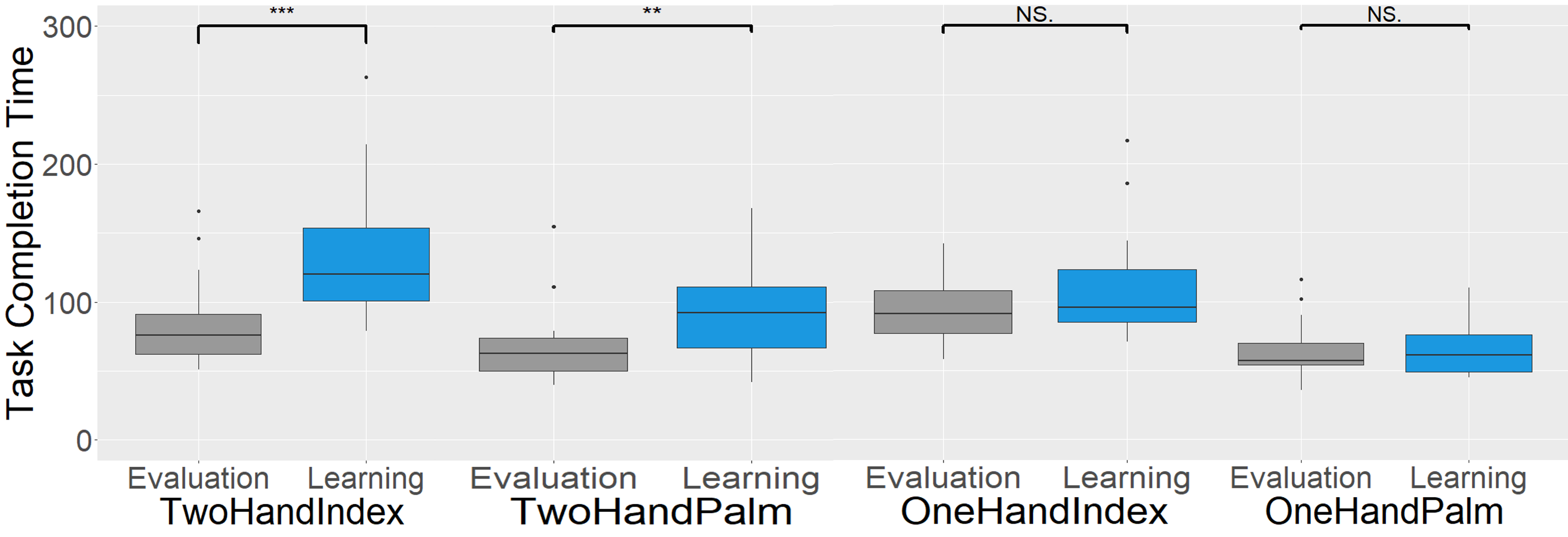}
\caption{Boxplot comparing task completion time between learning and evaluation phase. The values represent the time users required to fulfill the given task in seconds. Significance Levels: *** = 0.001; ** = 0.01; * = 0.05; NS = No Significance.} \label{fig:TCComparison}
\end{figure}

To answer this question, the task completion time was taken into account. Levene's test was conducted in order to ensure homogeneity of the input data ($p > 0.05$). One-way ANOVA was used in order to answer whether users are faster at completing the given task after performing a training. A comparison was made between the learning phase of each technique and the corresponding evaluation phase. The average values are depicted in \autoref{fig:TCComparison}. The results of the ANOVAs are: TwoHandIndex: $F(1,40) = 12.38, p = 0.001$; TwoHandPalm: $F(1,40) = 8.298, p = 0.006$; OneHandIndex: $F(1,40) = 2.04, p = 0.161$; OneHandPalm: $F(1,40) = 0.144, p = 0.707$. \par 
The results showed significant difference in the task completion time for the techniques TwoHandIndex and TwoHandPalm with $p < 0.05$. The techniques OneHandIndex and OneHandPalm did not show significance with $p > 0.05$. Therefore, it can be concluded that users performed significantly faster after conducting a learning phase for the two-handed techniques. The one-handed techniques however did not show significant improvements.

\subsection{How much do lay people adapt to the limitations of a hand tracking device after a first trial session?}
\label{sec:q2}
\begin{figure}[h]
\includegraphics[width=\textwidth]{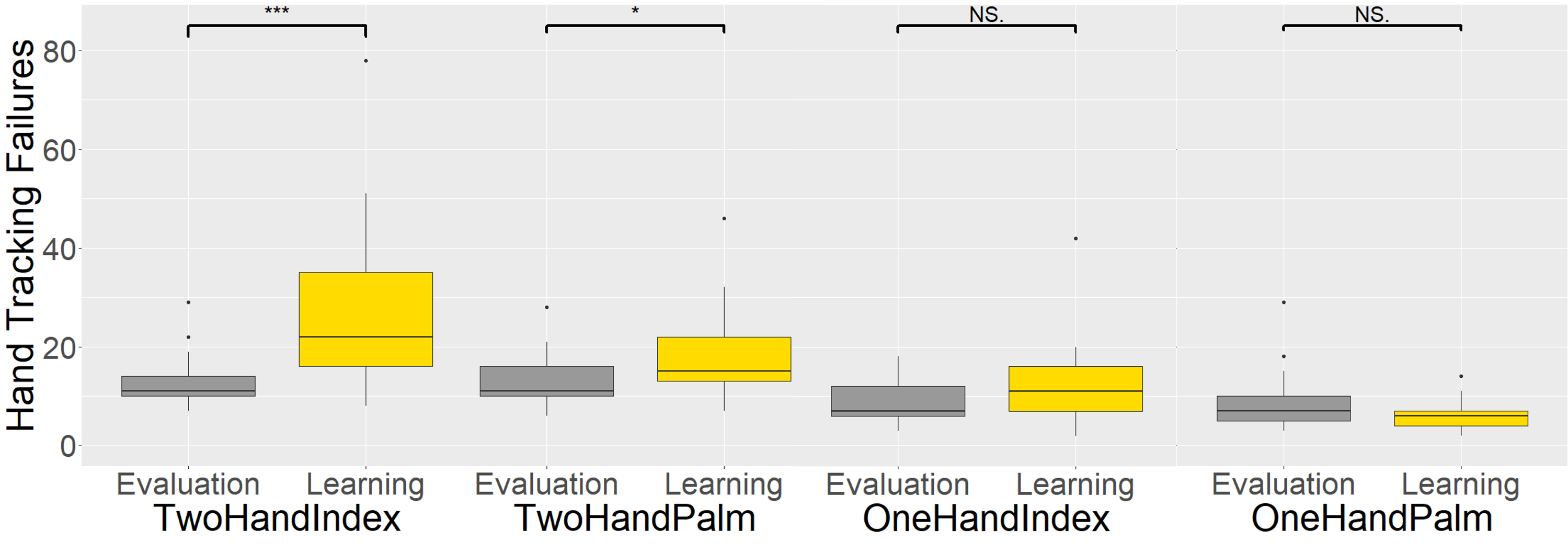}
\caption{Boxplot comparing the number of times the tracking has failed between learning and evaluation phase for each technique. Significance Levels: *** = 0.001; ** = 0.01; * = 0.05; NS = No Significance.} \label{fig:htlcomparison}
\end{figure}

Today, hand tracking devices have several limitations such as occlusion, low field of view, and tracking range. Scientists working in this field know these limitations and already avoid them unconsciously. Non-experts who have never been exposed to this technology will discover many of these limitations. This inevitably leads to many tracking errors until the user becomes aware of why the system has problems. For this reason, the number of times the hand tracking failed during a session was used as an indicator to answer this research question. Once the user's hands were no longer tracked, it was considered a loss of hand tracking. It can be said that users unconsciously and unintentionally move their hands out of the sensor's FOV because they are not accustomed to the technology. Therefore, this metric was used as an indicator of lay peoples learning effect of the chosen techniques. The average values are depicted in \autoref{fig:htlcomparison}. One-way ANOVA was used to find significant improvements between the learning and evaluation phase. Levene's test assured homogeneity of the input data. \par 
The result of the ANVOAs are:
TwoHandIndex: $F(1,40) = 12.69, p = 0.001$; 
TwoHandPalm: $F(1,40) = 5.727, p = 0.021$; 
OneHandIndex: $F(1,40) = 2.003, p = 0.165$; 
OneHandPalm: $F(1,40) = 1.898, p = 0.176$. \par

These results indicate, that the two-handed techniques showed significant improvements between the learning phase and the evaluation phase. The one-handed techniques did not show significance. The two-handed techniques show overall increased tracking errors compared to the one-handed techniques. Therefore, it can be concluded that people perform better when using one-handed techniques. However, users are also able to significantly improve with the two-handed techniques by only doing one prior session with the technique.

\subsection{Will lay people significantly improve their efficiency (less number of teleportations) when using hand gesture based locomotion during a second trial session?}
\begin{figure}[h]
\includegraphics[width=\textwidth]{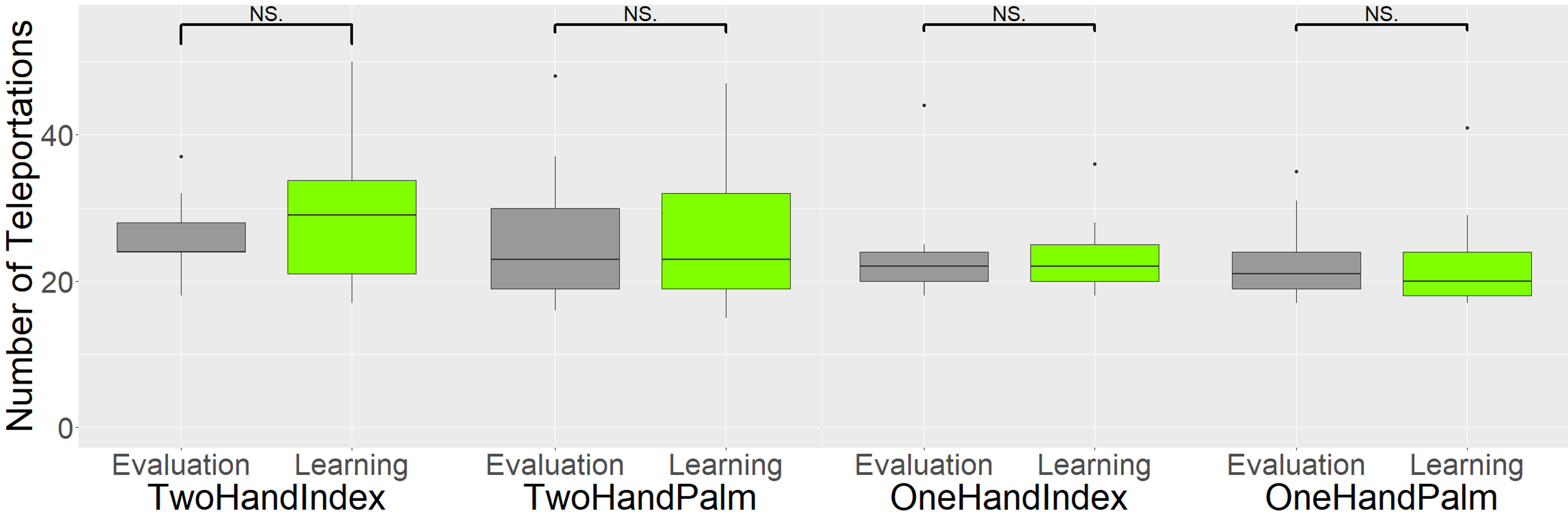}
\caption{Boxplot comparing the number of teleportations required to reach the goal between learning and evaluation phase. Significance Levels: *** = 0.001; ** = 0.01; * = 0.05; NS = No Significance.} \label{fig:NTComparison}
\end{figure}

To answer this question, the number of teleportations was considered. Levene's test was conducted in order to ensure homogeneity of the input data. One-way ANOVA was used to identify significant differences between the learning and evaluation phase of the experiment. The learning phase of each technique was compared with the corresponding evaluation phase. The average values are depicted in Figure \ref{fig:NTComparison}. The result of the ANOVAs are: 
TwoHandIndex: $F(1,40) = 3.214, p = 0.08$; 
TwoHandPalm: $F(1,40) = 0.001, p = 0.971$; 
OneHandIndex: $F(1,40) = 0.001, p = 0.975$; 
OneHandPalm: $F(1,40) = 0.008, p = 0.931$. \par
According to the one-way ANOVAs, there was no significant improvement observed for individual techniques between learning and evaluation phase ($p > 0.05$). Therefore, no learning effect could be observed in teleportation behavior. 

\section{Discussion}
\label{sec:discussion}
The participants showed a significant improvement in task completion time for the two-handed techniques. The use of one-handed techniques showed no significant improvement between the first and second phase of the experiment. During the experiment, subjects already expressed that one-handed techniques seem to be more intuitive and henceforth would explain these results. However, after performing the two-handed techniques for a second time, there was already significant improvement noticeable. This result is also backed by the fact that users significantly improved in regards to hand tracking failures. 
In the first phase, the users were uneasy because they first had to understand the limitations of the hand tracker. In the second phase, a clear improvement was noticeable. The number of teleportations required to reach the goal did not significantly vary between learning and evaluation phase. This could mean that the users understood how to achieve the goal and the task was straightforward to understand. \par


\section{Conclusion}
\label{sec:conclusion}
This paper investigated the learning effect of lay people performing hand gesture-based locomotion. A user study with 21 participants was conducted. In this study, four locomotion methods were utilised and the experiment was divided into a learning phase and an evaluation phase. All four methods were carried out twice by the subjects. The first time a method was performed was referred to as the learning phase and the second time as the evaluation phase. The study revealed significant improvements for the subjects while using two-handed techniques. The participants were considerably faster and significantly improved at using the hand tracking device. Therefore, it can be said that users struggle at first and then, with just one more trial run, they can significantly adapt to gesture-based systems with two hands. Furthermore, no significant learning effect was observed using one-handed techniques.  \par 
\subsubsection*{Acknowledgements}
\label{sec:acks}
Part of this work was funded by the Bundesministerium für Bildung und Forschung (BMBF) in the context of ODPfalz under Grant 03IHS075B. This work was also supported by the EU Research and Innovation programme Horizon 2020 (project INFINITY) under the grant agreement ID: 883293.
%
%
%
%
%
%
\bibliographystyle{splncs04}
\bibliography{bibo}

\end{document}